\documentclass[12pt,draftclsnofoot,journal,letterpaper,onecolumn]{IEEEtran}

\usepackage[T1]{fontenc}
\usepackage{cite}
\ifCLASSINFOpdf
  \usepackage[pdftex]{graphicx}
\else
  \usepackage[dvipdfmx]{graphicx}
\fi
\usepackage[dvipdfmx]{hyperref}
\usepackage{amsmath}
\usepackage[cmintegrals]{newtxmath}
\usepackage{bm}
\usepackage{flushend}
\usepackage{color}
\usepackage{url}
\usepackage{subfig}

\newcommand{\B}{\mathcal{B}}\newcommand{\F}{\mathcal{F}}
\newcommand{\K}{\mathcal{K}}
\newcommand{\Prb}{\mathbb{P}}\newcommand{\Exp}{\mathbb{E}}
\newcommand{\Var}{\mathsf{Var}}\newcommand{\Cov}{\mathsf{Cov}}
\newcommand{\CV}{\mathsf{CV}}\newcommand{\CC}{\mathsf{CC}}
\newcommand{\N}{\mathbb{N}}\newcommand{\R}{\mathbb{R}}
\newcommand{\Z}{\mathbb{Z}}
\newcommand{\dd}{\mathrm{d}}
\newcommand{\Lpl}{\mathcal{L}}
\newcommand{\bsym}[1]{\boldsymbol{#1}}
\newcommand{\ind}[1]{\boldsymbol{1}_{#1}}
\let\Bar\overline\let\UBar\underline

\newtheorem{assumption}{Assumption}
\newtheorem{proposition}{Proposition}
\newtheorem{theorem}{Theorem}
\newtheorem{corollary}{Corollary}
\newtheorem{lemma}{Lemma}
\newtheorem{remark}{Remark}

\title{Joint Performance Analysis of Ages of Information in a
  Multi-source Pushout Server}

\author{Yukang Jiang and Naoto Miyoshi
  \thanks{Y. Jiang and N. Miyoshi are with Department of Mathematical
    and Computing Science, Tokyo Institute of Technology, Tokyo,
    Japan. E-mail: miyoshi@is.titech.ac.jp.}
  \thanks{The support of the Japan Society for the Promotion of
    Science~(JSPS) Grant-in-Aid for Scientific Research (C) 19K11838
    is gratefully acknowledged.}
}

\begin{document}\sloppy\allowbreak\allowdisplaybreaks
\maketitle

\begin{abstract}
Age of information~(AoI) has been widely accepted as a measure
quantifying freshness of status information in real-time status update
systems.
In many of such systems, multiple sources share a limited network
resource and therefore the AoIs defined for the individual sources
should be correlated with each other.
However, there are not found any results studying the correlation of
two or more AoIs in a status update system with multiple sources.
In this work, we consider a multi-source system sharing a common service
facility and provide a framework to investigate joint performance of
the multiple AoIs.
We then apply our framework to a simple pushout server with multiple
sources and derive a closed-form formula of the joint Laplace
transform of the AoIs in the case with independent M/G inputs.
We further show some properties of the correlation coefficient of
AoIs in the two-source system.
\end{abstract}


\section{Introduction}\label{sec:Intro}

Freshness of status information is crucial in real-time status update
systems seen, for example, in weather reports, autonomous driving,
stock market trading and so on.
\textit{Age of information~(AoI)\/} has been widely accepted in this
decade as a measure quantifying the freshness of information in status
update systems where information sources transmit packets containing
status updates to destination monitors through a communication
network.
Specifically, the AoI is defined as the elapsed time since the
information currently displayed on a monitor is generated and
timestamped at the source.
In many of such systems, multiple sources share a limited network
resource and therefore the AoIs defined for the individual sources
should be correlated with each other.
However, to the best of the knowledge of the authors, there are no
results studying the correlation of two or more AoIs in a status
update system with multiple sources (except for the first author's
preliminary work~\cite{JianTokuWadaYaji20}).
To investigate the correlation is essential when, for example, we
consider a nonlinear penalty function of multiple AoIs.
In this work, we consider a multi-source system sharing a common service
facility and provide a framework to investigate joint performance of
the AoIs defined for the individual sources.

\subsection{Related work}
Since the advent in \cite{KaulGrutRaiKenn11,KaulYateGrut12a}, a large
amount of literature has emerged on development of the AoI concept due
to its importance and availability in a wide range of information
communication systems.
A complete review falls out of the reach of this paper and we only
highlight some relevant results to ours.
Interested readers are referred to recent
monographs~\cite{KostPappAnge17,SunKadoTalaModiSrik19} and references
therein.

The AoI was first introduced in \cite{KaulGrutRaiKenn11} in the
context of vehicular networks and queueing-theoretic technique was
applied in \cite{KaulYateGrut12a} to analyze the mean AoI under the
ergodicity assumption.
These results were then extended to various multi-source systems
in~\cite{KaulYateGrut12,YateKaul12,YateKaul19}.
A more tractable metric, \textit{peak AoI~(PAoI)}---the AoI
immediately before an update, was introduced in
\cite{CostCodrEphr14,CostCodrEphr16}, which characterized not only the
mean but also the probability distribution of the PAoI for various
queueing systems under the ergodicity.
Not only the mean, expected nonlinear functions of the AoI were
examined in~\cite{KostPappEphrAnge17,KostPappEphrAnge20}.

In the early stage of the development, they assumed that time
intervals of packet generations and their service times are either
exponentially distributed or deterministic; that is, they considered
M/M, M/D or D/M inputs in the queueing notation.
Recently, some researchers have challenged to incorporate more general
probability distributions.
In~\cite{NajmNass16}, gamma distributed service times were assumed
for Poisson arrival systems and \cite{HuanModi15,NajmTela18} considered
more general service time distributions in multi-source systems with
independent Poisson arrivals; that is, they treated independent M/G
input processes.
Furthermore,
\cite{InouMasuTakiTana17,InouMasuTakiTana19,KesiKonsZaza19}
studied more general frameworks of packet arrival and service time
processes and derived general formulas satisfied by the stationary
distribution and its Laplace transform of the AoI, where
\cite{InouMasuTakiTana17,InouMasuTakiTana19} adopted the technique of
sample path analysis while \cite{KesiKonsZaza19} worked based on the
Palm calculus within the stationary framework (see, e.g.,
\cite{BaccBrem03} for the Palm calculus).
However, there are no results confronting joint performance of
multiple AoIs in multi-source systems within general frameworks.

\subsection{Contribution}

In this paper, we consider a multi-source system sharing a service
facility as in
\cite{KaulYateGrut12,YateKaul12,YateKaul19,HuanModi15,NajmTela18} and
provide a framework to investigate joint performance of the AoIs
defined for the individual sources.
We first consider a general multi-source system, where the time
sequence representing service completions (and status updates) follows
a stationary point process on the real line, and derive a formula
satisfied by the joint Laplace transform of the stationary AoIs.
A tool for our analysis is the Palm calculus within the stationary
framework as in~\cite{KesiKonsZaza19}, where we do not require the
ergodic assumption but, once the ergodicity is assumed, we can obtain
the same results as those from the sample path analysis by the ergodic
theorem.
Our formula is so general and is applicable to many multi-source
systems.
We then apply this formula to a simple pushout server, where the
system has a single server and each generated packet is immediately
started for service without waiting; that is, the ongoing service of
another packet (if any) is interrupted and replaced by the new one.
In the case with independent M/G input processes, we derive a
closed-form formula of the joint Laplace transform of the AoIs.
Furthermore, we reveal some properties of the correlation coefficient
of the AoIs in the two-source system.


\subsection{Organization}

The rest of the paper is organized as follows.
In the next section, we describe a general multi-source system and
derive a formula satisfied by the joint Laplace transform of the
multiple AoIs.
We also provide a formula for a single-source system, which
corresponds to the review of the results
in~\cite{InouMasuTakiTana17,InouMasuTakiTana19} within our stationary
framework.
In Section~\ref{sec:Pushout}, we apply our formula to a multi-source
pushout server.
We first confirm in~\ref{subsec:pushout} that our multi-source pushout
server is definitely within our framework  and then derive a
closed-form formula of the joint Laplace transform of the AoIs in the
case with independent M/G inputs in~\ref{subsec:M/G/1/1}.
Some properties of the correlation coefficient of the AoIs in the
two-source system are revealed in~\ref{subsec:cc}.
These properties of the correlation coefficient are confirmed through
numerical experiments in Section~\ref{sec:experiments}.
Finally, concluding remarks and future work are discussed in
Section~\ref{sec:conclusion}.

\section{General multi-source system}

In this section, we consider a general multi-source single-server
system.
There are $K$~($\in\N=\{1,2,\ldots\}$) sources generating packets with
different kinds of status information and each source has its
dedicated monitor.
The set of the sources is denoted by $\K = \{1,2,\ldots,K\}$.
The system has a single server and, after a packet from
source~$k\in\K$ is processed by the server, the status information in
the packet is immediately displayed on monitor~$k$.
We note that not all packets are completed for service and some may be
discarded and lost (due to buffer overflows, packet deadlines or
pushouts).
We, for the moment, ignore such lost packets and focus on those being
completed for service.

Let $\Psi$ denote a point process on $\R=(-\infty,\infty)$ counting
the times at each of which the service of a packet is completed and
the status information on one of the monitors is updated, and let
$\{U_n\}_{n\in\Z}$ denote the corresponding time sequence, where
$\Z=\{\cdots,-1,0,1,2,\ldots\}$.
For $n\in\Z$, let $C_n$ and $D_n$ denote respectively the source and
the delay of the packet whose service is completed at $U_n$; that is,
the status information updated at $U_n$ is generated and timestamped
at $U_n-D_n$ by source~$C_n$.
For each $k\in\K$, let $\Psi_k$ denote a sub-process of $\Psi$
counting the times of service completions of source~$k$ packets; that
is,
\begin{equation}\label{eq:sub-departure}
  \Psi_k(B) = \sum_{n\in\Z}\ind{\{k\}}(C_n)\ind{B}(U_n),
  \quad B\in\B(\R),
\end{equation}
where $\ind{E}$ denotes the indicator function of a set $E$.
Clearly, $\Psi_k$, $k\in\K$, are mutually disjoint and satisfy
$\Psi=\sum_{k=1}^K\Psi_k$.
We impose the following assumption on the marked point
process~$\Psi_{C,D}$ corresponding to $\{(U_n,C_n,D_n)\}_{n\in\Z}$.

\begin{assumption}\label{asm1}\begin{enumerate}
\item\label{asm11} The point process~$\Psi$ is simple almost surely in
  a probability~$\Prb$ ($\Prb$-a.s.), where the rule of subscripts is
  such that $\cdots<U_{-1}<U_0\le0<U_1<\cdots$ conventionally (see,
  e.g., \cite{BaccBrem03}).
\item\label{asm12} The marked point process $\Psi_{C,D}$ is stationary
  in $\Prb$ with mark space $\K\times[0,\infty)$.
\item\label{asm13} The point process~$\Psi$ has positive and finite
  intensity~$\lambda_\Psi = \Exp[\Psi(0,1]]$, where $\Exp$   denotes
  the expectation with respect to $\Prb$.
\item\label{asm14} $\Prb_\Psi^0(C_0=k)>0$ for all $k\in\K$, where
  $\Prb_\Psi^0$ denotes the Palm probability for $\Psi$.
\item\label{asm15} $\Prb_{\Psi_k}^0(D_0<\infty)=1$ for all $k\in\K$,
  where $\Prb_{\Psi_k}^0$ denotes the Palm probability for $\Psi_k$.
\end{enumerate}
\end{assumption}

Note that the Palm probability $\Prb_\Psi^0$ is well defined under
Assumptions~\ref{asm1}-\ref{asm12}) and~\ref{asm13}) (see
\cite{BaccBrem03}) and that $\Prb_\Psi^0(U_0=0)=1$.
Furthermore, under Assumption~\ref{asm1}-\ref{asm12}),
$\{(C_n,D_n)\}_{n\in\Z}$ is stationary in $\Prb_\Psi^0$.
Assumption~\ref{asm1}-\ref{asm14}) does not restrict us since we can
redefine $\K$ by $\K\setminus\{k\}$ if $\Prb_\Psi^0(C_0=k)=0$ for some
$k\in\K$.
The intensity~$\lambda_{\Psi_k}$ of the sub-process~$\Psi_k$ is given
by $\lambda_{\Psi_k} = \lambda_\Psi\,\Prb_\Psi^0(C_0=k)$ for each
$k\in\K$, so that the Palm probability~$\Prb_{\Psi_k}^0$ is also well
defined under \ref{asm1}-\ref{asm12})--\ref{asm14}) and satisfies
$\Prb_{\Psi_k}^0(\cdot) = \Prb_\Psi^0(\cdot \mid C_0 = k)$.
Let $\{U_{k,n}\}_{n\in\Z}$, $k\in\K$, denote the time sequence
corresponding to $\Psi_k$ satisfying
$\cdots<U_{k,-1}<U_{k,0}\le0<U_{k,1}<\cdots$, and let also $D_{k,n}$
denote the delay of the source~$k$ packet whose service is completed
at $U_{k,n}$ (note that $D_{k,0}=D_0$ $\Prb_{\Psi_k}^0$-a.s.).
Then, the AoI process~$\{A_k(t)\}_{t\in\R}$ for source~$k\in\K$ is
defined as (see \cite{YateKaul12,YateKaul19} for AoI in multi-source
systems)
\begin{equation}\label{eq:AoI}
  A_k(t) = D_{k,n} + t - U_{k,n},
  \quad t\in[U_{k,n}, U_{k,n+1}),\; n\in\Z.
\end{equation}
This definition indicates that the AoI of source~$k$ represents the
elapsed time since the information currently displayed on monitor~$k$
is generated and timestamped; that is, it is set to the delay of a
source~$k$ packet at its service completion time and increases
linearly until the status information from source~$k$ is next updated.

\begin{lemma}\label{lem:joint-stationary}
For each $k\in\K$, the marked point process $\Psi_{k,D}$ corresponding
to $\{(U_{k,n},\, D_{k,n})\}_{n\in\Z}$ and also the AoI processes
$\{A_k(t)\}_{t\in\R}$ are jointly stationary with the marked point
process $\Psi_{C,D}$ under Assumption \ref{asm1}.
Furthermore, $A_k(0)$ is $\Prb$-a.s.\ finite under the same
assumption.
\end{lemma}

\begin{IEEEproof}
The proof relies on a technical discussion within the stationary
framework and is given in Appendix~\ref{sec:Prf_stationary}.
\end{IEEEproof}

While the AoI just after an update is equal to the delay
$D_{k,n}=A(U_{k,n})$, that just before an update is called the PAoI
(see \cite{CostCodrEphr14,CostCodrEphr16}); that is, for each
$k\in\K$, the sequence $\{P_{k,n}\}_{n\in\Z}$ of PAoIs is defined as
$P_{k,n} = A(U_{k,n}-)$, $n\in\Z$, and is also stationary in
$\Prb_{\Psi_k}^0$.
Prior to the joint performance analysis of $A_1(t),\ldots,A_K(t)$,
$t\in\R$, we review the results of
\cite{InouMasuTakiTana17,InouMasuTakiTana19} for a single-source 
single-server system within our stationary framework, which is also
useful in the marginal performance analysis of multi-source systems.

\begin{proposition}[Cf.\ \cite{InouMasuTakiTana17,InouMasuTakiTana19}]\label{prp:AoI}
Consider a single-source system satisfying Assumption~\ref{asm1} with
$K=1$, where $A(t)=A_1(t)$, $D_n=D_{1,n}$ and $P_n=P_{1,n}$ in
\eqref{eq:AoI}.
Then, the stationary distribution of the AoI satisfies
\begin{equation}\label{eq:AoI_dist}
  \Prb(A(0)\le x)
  = \lambda_\Psi
    \int_0^x
      \bigl(\Prb_\Psi^0(P_0>u) - \Prb_\Psi^0(D_0>u)\bigr)\,
    \dd u,
  \quad x\ge0.
\end{equation}
Let $\Lpl_A(s) = \Exp[e^{-s A(0)}]$, $\Lpl_D(s) = \Exp_\Psi^0[e^{-s
    D_0}]$ and $\Lpl_P(s) = \Exp_\Psi^0[e^{-s P_0}]$, $s\in\R$, denote
the Laplace transforms of $A(0)$, $D_0$ and $P_0$, respectively, where
$\Exp_\Psi^0$ denotes the expectation with respect to the Palm
probability $\Prb_\Psi^0$.
Then, the following relation holds;
\begin{equation}\label{eq:AoI_Lpl}
  s\,\Lpl_A(s)
  = \lambda_\Psi\,\bigl(\Lpl_D(s) - \Lpl_P(s)\bigr),
\end{equation}
for $s\in\R$ such that the Laplace transforms on both the sides exist.
\end{proposition}

\begin{IEEEproof}
Applying the Palm inversion formula (see \cite[p.~20]{BaccBrem03}),
\begin{align*}\label{eq:AoI-D}
  \Prb(A(0)\le x)
  &= \lambda_\Psi\,\Exp_\Psi^0\biggl[
       \int_{[0,\,U_1)}
         \ind{[0,\,x]}(A(t))\,
       \dd t
     \biggr]
  \\
  &= \lambda_\Psi\,\Exp_\Psi^0\biggl[
       \int_{[D_0,\,P_1)}
         \ind{[0,\,x]}(u)\,
       \dd u
     \biggr]
  \nonumber\\
  &= \lambda_\Psi\,\Exp_\Psi^0[ P_1\wedge x - D_0\wedge x],
  \quad x\ge0,
\end{align*}
where $a\wedge b=\min(a,b)$ for $a,b\in\R$ and the second equality
follows from \eqref{eq:AoI}; that is, $A(t)$ increases linearly from
$D_0$ to $P_1$ for $t\in[U_0, U_1)$.
The last expression above immediately derives \eqref{eq:AoI_dist}
since $P_1$ is distributionally equal to $P_0$ in $\Prb_\Psi^0$.
The formula~\eqref{eq:AoI_Lpl} is obtained from \eqref{eq:AoI_dist} as
\begin{align*}
  \Lpl_A(s)
  &= \int_0^\infty e^{-s x}\,\Prb(A(0)\in\dd x)
  \\
  &= \lambda_\Psi\int_0^\infty
       e^{-s x}\,\bigl(\Prb_\Psi^0(P_0>x) - \Prb_\Psi^0(D_0>x)\bigr)\,
     \dd x,
\end{align*}
using $\Exp[e^{-s X}] = 1 - s\int_0^\infty e^{-s x}\,\Prb(X>x)\,\dd x$
for a random variable~$X$ and $s\in\R$ such that $\Exp[e^{-s X}]$ is
finite.
\end{IEEEproof}

\begin{remark}
Formulas~\eqref{eq:AoI_dist} and \eqref{eq:AoI_Lpl} respectively
correspond to the results of Theorem~14~(i) and~(ii) in
\cite{InouMasuTakiTana19}.
As opposed to \cite{InouMasuTakiTana17,InouMasuTakiTana19}, however,
Proposition~\ref{prp:AoI} does not require the ergodicity of $\{(D_n,
P_n)\}_{n\in\Z}$.
These formulas suggest that we can obtain the stationary distribution
of the AoI and the corresponding Laplace transform once the stationary
distributions of the delay and PAoI are available.
The formula~\eqref{eq:AoI_dist} also implies that the stationary
distribution of the AoI has the density
function~$\lambda_\Psi\,\bigl(\Prb_\Psi^0(P_0>x) -
\Prb_\Psi^0(D_0>x)\bigr)$, $x\ge0$.
Taking $x\to\infty$ in \eqref{eq:AoI_dist}, we have $\Exp_\Psi^0[P_0]
= \Exp_\Psi^0[D_0] + 1/\lambda_\Psi$, which intuitively makes sense
since a PAoI is the sum of a delay and its subsequent interdeparture
time (see \eqref{eq:AoI}).
\end{remark}

Proposition~\ref{prp:AoI} can be applied to the marginal distribution
of each AoI in a system with $K\ge2$.

\begin{corollary}\label{crl:marginalAoI}
For a system with $K$ sources satisfying Assumption~\ref{asm1}, the
marginal stationary distribution of the AoI for source~$k\in\K$
satisfies
\begin{equation}\label{eq:AoI_marginal_dist}
  \Prb(A_k(0)\le x)
  = \lambda_{\Psi_k}
    \int_0^x
      \bigl(\Prb_{\Psi_k}^0(P_{k,0}>u) - \Prb_{\Psi_k}^0(D_{k,0}>u)\bigr)\,
    \dd u,
  \quad x\ge0.
\end{equation}  
Let $\Lpl_{A_k}(s) = \Exp[e^{-s A_k(0)}]$, $\Lpl_{D_k}(s) =
\Exp_{\Psi_k}^0[e^{-s D_{k,0}}]$ and $\Lpl_{P_k}(s) =
\Exp_{\Psi_k}^0[e^{-s P_{k,0}}]$, $s\in\R$, denote the Laplace
transforms of $A_k(0)$, $D_{k,0}$ and $P_{k,0}$, respectively, where
$\Exp_{\Psi_k}^0$ denotes the expectation with respect to
$\Prb_{\Psi_k}^0$.
Then, the following relation holds;
\begin{equation}\label{eq:AoI_marginal_Lpl}
  s\,\Lpl_{A_k}(s)
  = \lambda_{\Psi_k}\,
    \bigl(\Lpl_{D_k}(s) - \Lpl_{P_k}(s)\bigr),
\end{equation}
for $s\in\R$ such that the Laplace transforms on both the sides exist.
\end{corollary}

Next, we consider the joint performance of multiple AoIs.
Let $\Lpl_{\bsym{A}}$ on $\R^K$ denote the joint Laplace transform of
$A_1(0),\ldots,A_K(0)$; that is,
\begin{equation}\label{eq:JointLpl}
  \Lpl_{\bsym{A}}(\bsym{s})
  = \Exp\biggl[
      \exp\biggl(-\sum_{k=1}^K s_k\,A_k(0)\biggr)
  \biggr],
  \quad \bsym{s} = (s_1,\ldots,s_K)\in\R^K.
\end{equation}
Note that $\Lpl_{\bsym{A}}$ does not always exist on the whole space
of $\R^K$.
We derive a general formula satisfied by $\Lpl_{\bsym{A}}$ as far as
it exists, which is applicable to the analysis of many multi-source
single-server systems satisfying Assumption~\ref{asm1}.
Let $(\eta_1,\ldots,\eta_K)$ denote a random permutation of $\K$
satisfying
\begin{equation}\label{eq:eta}
  U_0 = U_{\eta_1,\,0} > U_{\eta_2,\,0} > \cdots > U_{\eta_K,\,0}.
\end{equation}
That is, $\eta_j$ represents the source such that the status
information on its monitor at time~$0$ is the $j$th newest among $\K$
(the information on monitor~$\eta_1$ is most recently updated while
monitor~$\eta_K$ displays the oldest information).
Note that $U_{\eta_1,\,0}, \ldots, U_{\eta_K,\,0}$ satisfy the
relation;
\begin{equation}\label{eq:U_eta}
  -U_{\eta_k,\,0}
  = -U_{\eta_1,\,0} + \sum_{j=2}^k(U_{\eta_{j-1},\,0} - U_{\eta_j,\,0}),
  \quad k=2,\ldots,K.
\end{equation}
In the following, we write $\Bar{s}_H = \sum_{i\in H}s_i$ for a
nonempty subset~$H\subset\K$ with $\Bar{s} = \Bar{s}_{\K} =
\sum_{i=1}^K s_i$ and $\eta[k] = \{\eta_k,\eta_{k+1},\ldots,\eta_K\}$
given the random permutation~$(\eta_1,\ldots,\eta_K)$
satisfying~\eqref{eq:eta}.

\begin{theorem}\label{thm:AoI_joint_Lpl}
For a $K$-source single-server system satisfying
Assumption~\ref{asm1}, the joint Laplace transform $\Lpl_{\bsym{A}}$
of the stationary AoIs~$A_1(0), \ldots, A_K(0)$ satisfies
\begin{equation}\label{eq:AoI_joint_Lpl}
  \Bar{s}\,\Lpl_{\bsym{A}}(\bsym{s})
  = \lambda_\Psi\,
    \Exp_\Psi^0\biggl[
      \bigl(1 - e^{-\Bar{s} U_1}\bigr)
      \prod_{k=1}^{K-1}
        \exp\Bigl\{
          -s_{\eta_k} D_{\eta_k,\,0}
          - \Bar{s}_{\eta[k+1]} (U_{\eta_k,\,0}-U_{\eta_{k+1},\,0})
        \Bigr\}
      \exp\bigl(-s_{\eta_K} D_{\eta_K,\,0}\bigr)
    \biggr],
\end{equation}
for $\bsym{s} = (s_1,\ldots,s_K) \in \R^K$ such that the expectation
on the right-hand side exists.
\end{theorem}

\begin{IEEEproof}
Applying the Palm inversion formula to \eqref{eq:JointLpl} and then
using \eqref{eq:AoI}, we have
\begin{align}\label{eq:multi-AoI-1}
  \Lpl_{\bsym{A}}(\bsym{s})
  &= \lambda_\Psi\,
     \Exp_\Psi^0\biggl[
       \int_{[0,\,U_1)}
         \exp\biggl(-\sum_{k=1}^K s_k A_k(t)\biggr)\,\dd t
     \biggr]
  \nonumber\\
  &= \lambda_\Psi\,
     \Exp_\Psi^0\biggl[
       \prod_{k=1}^K e^{-s_k (D_{k,0}-U_{k,0})}
       \int_{[0,\,U_1)}
         e^{- \Bar{s} t}\,\dd t
     \biggr].
\end{align}
It is immediate that $\Bar{s}\int_{[0,\,U_1)} e^{- \Bar{s} t}\,\dd t =
  1 - e^{-\Bar{s} U_1}$.
Furthermore, the relation~\eqref{eq:U_eta} on the event $\{U_0 =
U_{\eta_1,\,0} = 0\}$ implies that
\begin{align*}
  \prod_{k=1}^K e^{-s_k (D_{k,0}-U_{k,0})}
  &= \prod_{k=1}^K
       \exp\bigl(-s_{\eta_k}\,(D_{\eta_k,\,0}-U_{\eta_k,\,0})\bigr)
  \\
  &= \prod_{k=1}^K \exp\bigl(-s_{\eta_k} D_{\eta_k,\,0}\bigr)
     \prod_{k=2}^K\prod_{j=2}^k
       \exp\bigl(-s_{\eta_k} (U_{\eta_{j-1},\,0}-U_{\eta_j,\,0})\bigr)
  \\
  &= \prod_{k=1}^K \exp\bigl(-s_{\eta_k} D_{\eta_k,\,0}\bigr)
     \prod_{j=2}^K \exp\bigl(
       -\Bar{s}_{\eta[j]} (U_{\eta_{j-1},\,0}-U_{\eta_j,\,0})
     \bigr).
\end{align*}
Plugging this into \eqref{eq:multi-AoI-1} derives
\eqref{eq:AoI_joint_Lpl}.
\end{IEEEproof}

\begin{remark}
We can easily confirm that \eqref{eq:AoI_joint_Lpl} agrees with
\eqref{eq:AoI_Lpl} in Proposition~\ref{prp:AoI} when $K=1$
($\prod_{k=1}^0\cdot=1$ in this case) since $U_0=0$ and $P_1 =
D_0+U_1$ $\Prb_\Psi^0$-a.s.
On the other hand, it is not so straightforward to show that
\eqref{eq:AoI_joint_Lpl} with $s_j=0$ for all $j\in\K\setminus\{k\}$
agrees with \eqref{eq:AoI_marginal_Lpl} in
Corollary~\ref{crl:marginalAoI} because \eqref{eq:AoI_joint_Lpl} is
based on the Palm inversion formula with respect to $\Prb_\Psi$ (with
the integral on $[U_0,U_1)$) while \eqref{eq:AoI_marginal_Lpl} is on
that with respect to $\Prb_{\Psi_k}$ (with the integral on
$[U_{k,0},U_{k,1})$).
Therefore, Corollary~\ref{crl:marginalAoI} still makes sense in
marginal analysis of multi-source systems.
Note that the terms in the product in \eqref{eq:AoI_joint_Lpl} are
evaluated in mutually disjoint intervals~$(U_{\eta_K,\,0}-D_{\eta_K,\,0},
U_{\eta_K,\,0}], (U_{\eta_K,\,0}, U_{\eta_{K-1},\,0}], \ldots,
(U_{\eta_2,\,0},0]$ and $(0,U_1]$ on the event $\{U_0=0\}$.
This property can make formula \eqref{eq:AoI_joint_Lpl} useful for
analysis of a class of multi-source systems such that the sequence of
service completion times forms a regenerative process or an embedded
Markov chain.
\end{remark}

\section{Application to a multi-source pushout server}\label{sec:Pushout}

In this section, we apply the results in the preceding section to a
pushout server with $K$ sources.
We first confirm that the system satisfies Assumption~\ref{asm1} in
the preceding section and then derive a closed-form formula for the
joint Laplace transform of the AoIs in the case with independent M/G
input processes.
We further reveal some properties of the correlation coefficient of
the AoIs in the two-source system.

\subsection{Multi-source pushout server}\label{subsec:pushout}

We here describe the system consisting of a pushout server and $K$
sources with the dedicated monitors.
Let $\Phi$ denote a point process on $\R$ counting the times at each
of which a packet is generated and timestamped by any one of the
sources and let $\{T_n\}_{n\in\Z}$ denote the corresponding time
sequence.
For each $n\in\Z$, $c_n$ and $S_n$ denote respectively the source and
the required service time of the packet generated at $T_n$.
For each $k\in\K$, let $\Phi_k$ denote the sub-process of $\Phi$
counting the generation times of source~$k$ packets; that is,
\[
  \Phi_k(B) = \sum_{n\in\Z}\ind{\{k\}}(c_n)\ind{B}(T_n),
  \quad B\in\B(\R).
\]
We impose the following assumption on the marked point
process~$\Phi_{c,S}$ corresponding to $\{(T_n,c_n,S_n)\}_{n\in\Z}$.

\begin{assumption}\label{asm2}
\begin{enumerate}
\item\label{asm21} The point process~$\Phi$ is $\Prb$-a.s.\ simple,
  where $\{T_n\}_{n\in\Z}$ is numbered as
  $\cdots<T_{-1}<T_0\le0<T_1<\cdots$ conventionally.
\item\label{asm22} The marked point process $\Phi_{c,S}$ is stationary
  in $\Prb$ with mark space $\K\times[0,\infty)$.
\item\label{asm23} The point process~$\Phi$ has positive and finite
  intensity~$\lambda = \Exp[\Phi(0,1]]$.
\item\label{asm24} $\Prb_\Phi^0(c_0=k) > 0$ for all $k\in\K$, where
  $\Prb_\Phi^0$ denotes the Palm probability for $\Phi$.
\item\label{asm25} $\Prb_{\Phi_k}^0(S_0\le\tau_0)>0$ for all $k\in\K$,
  where $\tau_n=T_{n+1}-T_n$, $n\in\Z$, and $\Prb_{\Phi_k}^0$ denotes
  the Palm probability for $\Phi_k$.
\end{enumerate}
\end{assumption}

The Palm probability~$\Prb_\Phi^0$ in
Assumption~\ref{asm2}-\ref{asm24}) is well defined under
\ref{asm2}-\ref{asm22}) and~\ref{asm23}) and so are $\Prb_{\Phi_k}^0$,
$k\in\K$, under \ref{asm2}-\ref{asm22})--\ref{asm24}) since the
intensity~$\lambda_k$ of $\Phi_k$ is given by $\lambda_k =
\lambda\,\Prb_\Phi^0(c_0=k) \in(0,\infty)$.
Note here that $\Prb_{\Phi_k}^0(\cdot) = \Prb_\Phi^0(\cdot \mid c_0=k)$
holds for $k\in\K$.

The system has a single server and each generated packet is
immediately started for service without waiting.
If another one is in service at the generation time of a packet, the
service is interrupted and replaced by the new one (the interrupted
packet is pushed out and lost).
There is no priority among the sources and the service of any packet
can be interrupted by the next generated one from the same or other
sources.
The probability that a packet generated at source~$k$ is completed for
service without interruption is then given by
$\Prb_{\Phi_k}^0(S_0\le\tau_0)$, which is positive for all $k\in\K$
under Assumption~\ref{asm2}-\ref{asm25}).
When the service for a packet is completed without interruption, the
status information carried by the packet is displayed on the monitor
dedicated to the source of that packet.
Then, the marked point process $\Psi_{C,D}$, representing the service
completions considered in the preceding section, is expressed in terms
of $\Phi_{c,S}$ as
\begin{equation}\label{eq:input-output}
  \Psi_{C,D}(B\times\{k\}\times E)
  = \sum_{n\in\Z}
      \ind{B}(T_n+S_n)\,\ind{E\cap[0,\tau_n]}(S_n)\,
      \ind{\{k\}}(c_n),
  \quad B\in\B(\R),\; k\in\K,\; E\in\B([0,\infty)).
\end{equation}

\begin{lemma}\label{lem:input-output}
When the input marked point process~$\Phi_{c,S}$ satisfies
Assumption~\ref{asm2}, then the output process~$\Psi_{C,D}$ satisfies
Assumption~\ref{asm1} with
\begin{align}
  \lambda_\Psi
  &= \lambda\,\Prb_\Phi^0(S_0\le\tau_0)
  \nonumber\\
  &= \sum_{k=1}^K \lambda_k\,\Prb_{\Phi_k}^0(S_0\le\tau_0),
  \label{eq:lambda_Psi}\\
  \Prb_\Psi^0(C_0=k)
  &= \Prb_\Phi^0(c_0=k \mid S_0 \le \tau_0)
  \nonumber\\
  &= \frac{\lambda_k\,\Prb_{\Phi_k}^0(S_0 \le \tau_0)}
          {\sum_{j=1}^K \lambda_j\,\Prb_{\Phi_j}^0(S_0 \le \tau_0)}.
  \label{eq:P_Psi(C0=k)}
\end{align}
\end{lemma}  

\begin{IEEEproof}
The proof is also based on the stationary framework and is given in
Appendix~\ref{sec:Prf_input-output}.
\end{IEEEproof}

\subsection{Multi-source M/G/1/1 pushout server}\label{subsec:M/G/1/1}

In this subsection, by specifying the input point process as
independent homogeneous Poisson processes and assuming independence in
the service times, we derive a closed-form formula for the joint
Laplace transform of the AoIs~$A_1(0), \ldots, A_K(0)$.
We assume that $\Phi_1,\ldots,\Phi_K$ are mutually independent
homogeneous Poisson processes with positive and finite
intensities~$\lambda_1,\ldots,\lambda_K$.
The superposition theorem for Poisson processes~(see, e.g.,
\cite[p.~20]{LastPenr17}, \cite[p.~36]{BaccBlasKarr20}) then implies
that $\Phi =  \sum_{k=1}^K\Phi_k$ is also a homogeneous Poisson
process with intensity~$\lambda = \sum_{k=1}^K \lambda_k$.
We further assume that service times~$S_n$, $n\in\Z$, depend only on
their sources and, when the sources of packets are given, the service
times are mutually independent and independent of $\Phi_k$, $k\in\K$.
Namely, for any $m\in\N$, $n_1,\ldots,n_m\in\Z$,
$k_1,\ldots,k_m\in\K$, and $E_1,\ldots,E_m\in\B([0,\infty))$, we have
\[
  \Prb_\Phi^0\bigl(
    c_{n_1}=k_1, S_{n_1}\in E_1, \ldots, c_{n_m}=k_m, S_{n_m}\in E_m
  \bigr)
  = \frac{1}{\lambda^m}
     \prod_{j=1}^m
       \lambda_{k_j}\Prb_{\Phi_{k_j}}^0(S_0\in E_j).
\]
Let $\Lpl_{S,k}$ denote the Laplace transform of service times of
source~$k$ packets; that is, $\Lpl_{S,k}(s) = \Exp_{\Phi_k}^0[e^{-s
    S_0}]$, $s\in\R$ (it may be infinite for some $s<0$), where we
note that $\Exp_{\Phi_k}^0[\cdot] = \Exp_\Phi^0[\cdot\mid c_0=k]$ and
$\Exp_\Phi^0[\cdot] = \Exp[\cdot\mid T_0=0]$.
In this setup, Assumption~\ref{asm2} is satisfied and the probability
that a source~$k$ packet is completed for service without interruption
is given by
\begin{align*}
  \Prb_{\Phi_k}^0(S_0\le\tau_0)
  &= \Exp_{\Phi_k}^0\bigl[
       \Prb_{\Phi_k}^0(S_0\le\tau_0 \mid S_0)
     \bigr]
  \\  
  &= \Exp_{\Phi_k}^0[e^{-\lambda S_0}] = \Lpl_{S,k}(\lambda),
\end{align*}
where the second equality follows from the independent increments
property of a Poisson process.
Then, \eqref{eq:lambda_Psi} and \eqref{eq:P_Psi(C0=k)} in
Lemma~\ref{lem:input-output} are respectively rewritten as
$\lambda_\Psi = \lambda\,\Lpl_S(\lambda) = \sum_{k=1}^K
\lambda_k\,\Lpl_{S,k}(\lambda)$ and $\Prb_\Psi^0(C_0=k) =
\lambda_k\,\Lpl_{S,k}(\lambda)/\bigl(\lambda\,\Lpl_S(\lambda)\bigr)$
with $\Lpl_S(s) = \lambda^{-1}\sum_{k=1}^K \lambda_k\,\Lpl_{S,k}(s)$.
Furthermore, we use the following notation such that, for a nonempty
subset~$H\subset\K$,
\begin{align*}
  \Lpl_{S,H}(s)
  &= \Exp_\Phi^0[e^{-s S_0} \mid c_0\in H]
  \\
  &= \frac{1}{\Bar{\lambda}_H}
     \sum_{k\in H}\lambda_k\,\Lpl_{S,k}(s),
\end{align*}
with $\Bar{\lambda}_H = \sum_{k\in H}\lambda_k$.
Note that $\Lpl_{S,\K} = \Lpl_S$ and $\Lpl_{S,\{k\}} = \Lpl_{S,k}$ for
$k\in\K$.

First, we consider the marginal Laplace transform of the AoI for each
source.

\begin{proposition}
For the $K$-source M/G/1/1 pushout server described above, the
marginal Laplace transform~$\Lpl_{A_k}$ of the stationary AoI~$A_k(0)$
of source~$k$ is given by
\begin{equation}\label{eq:marginalAoI}
  \Lpl_{A_k}(s)
  = \frac{\lambda_k\,\Lpl_{S,k}(s+\lambda)}
         {s + \lambda_k\,\Lpl_{S,k}(s+\lambda)},
  \quad s\ge0,\; k\in\K.   
\end{equation}
\end{proposition}

\begin{IEEEproof}
We use \eqref{eq:AoI_marginal_Lpl} in Corollary~\ref{crl:marginalAoI}.
Note that $P_{k,1} = D_{k,0} + (U_{k,1}-U_{k,0})$ by \eqref{eq:AoI}
and the definition of the PAoI.
In our M/G/1/1 pushout server, since $D_{k,0}$ and $U_{k,1}-U_{k,0}$
are mutually independent and $U_{k,1}-U_{k,0}$ is also independent of
$\{C_0=k\}$ on the event $\{U_0=U_{k,0}=0\}$,
\eqref{eq:AoI_marginal_Lpl} is reduced to
\begin{equation}\label{eq:prf-marginalAoI0}
  s\,\Lpl_{A_k}(s)
  = \lambda_{\Psi_k}\,\Lpl_{D_k}(s)\,
    \bigl(1 - \Exp_\Psi^0[e^{-s U_{k,1}}]\bigr).
\end{equation}
First, Neveu's exchange formula (see \cite[p.~21]{BaccBrem03})
implies that
\begin{align}\label{eq:prf-marginalAoI1}
  \lambda_{\Psi_k}\,\Lpl_{D_k}(s)
  &= \lambda_k\,\Exp_{\Phi_k}^0\biggl[
       \sum_{n\in\Z}
         e^{- s D_{k,n}}\,
         \ind{[0,T_{k,1})}(U_{k,n}) 
     \biggr]
  \nonumber\\       
  &= \lambda_k\,\Exp_{\Phi_k}^0\bigl[
       e^{-s S_0}\,\ind{\{S_0 \le \tau_0\}}
     \bigr]
  \nonumber\\
  &= \lambda_k\,\Exp_{\Phi_k}^0\bigl[
       e^{-s S_0}\,\Prb_{\Phi_k}^0(S_0 \le \tau_0 \mid S_0)
     \bigr]
  \nonumber\\
  &= \lambda_k\,\Lpl_{S,k}(s+\lambda),
\end{align}
where $\{T_{k,n}\}_{n\in\Z}$ denotes the sub-sequence of
$\{T_n\}_{n\in\Z}$ corresponding to $\Phi_k$ satisfying $\cdots <
T_{k,0} \le 0 < T_{k,1} < \cdots$ and $\Prb_{\Phi_k}^0(T_{k,0}=0)=1$.
The second equality in \eqref{eq:prf-marginalAoI1} follows from the
observation that there exists at most one service completion of a
source~$k$ packet during $[T_{k,0},\, T_{k,1})$ and it occurs only
when the packet generated at $T_{k,0}$ is completed for service
without interruption.
  
Next, we consider $\Exp_\Psi^0[e^{-s U_{k,1}}]$ in
\eqref{eq:prf-marginalAoI0}.
Note that there may be one or more service completions during $(U_0,
U_{k,1})$, but if any, they must be of the sources in $\K\setminus\{k\}$.
Since $U_m = U_1 + \sum_{n=1}^{m-1} (U_{n+1}-U_n)$ for $m\ge1$ (where
$\sum_{n=1}^0\cdot=0$) and the server is always reset at $U_n$,
$n\in\Z$, we have
\begin{align}\label{eq:prf-marginalAoI2}
  \Exp_\Psi^0[e^{-s U_{k,1}}]
  &= \sum_{m=1}^\infty
       \Exp_\Psi^0\bigl[
         e^{-s U_{k,1}}\,\ind{\{U_{k,1} = U_m\}}
       \bigr]
  \nonumber\\
  &= \Exp_\Psi^0\bigl[e^{-s U_1}\,\ind{\{C_1=k\}}\bigr]
     \sum_{m=1}^\infty
       \bigl(
         \Exp_\Psi^0\bigl[
           e^{-s U_1}\,\ind{\{C_1 \in \K\setminus\{k\}\}}
         \bigr]
       \bigr)^{m-1}.
\end{align}
We solve $\Exp_\Psi^0\bigl[e^{-s U_1}\,\ind{\{C_1=k\}}\bigr]$ above.
Let $B_n$, $n\in\Z$, denote the time length of the busy period starting at
$T_n$ and ending at the next service completion.
Since $T_1$ is independent of the event $\{U_0=0\}$ due to the
independent increments property of a Poisson process and $B_n$ is
initialized at $T_n$, we have
\begin{align}\label{eq:prf-marginalAoI3}
  \Exp_\Psi^0\bigl[e^{-s U_1}\,\ind{\{C_1=k\}}\bigr]
  &= \Exp_\Psi^0\bigl[e^{-s (T_1 + B_1)}\,\ind{\{C_1=k\}}\bigr]
  \nonumber\\
  &= \Exp[e^{-s T_1}]\,
     \Exp_\Phi^0\bigl[e^{-s B_0}\,\ind{\{C_1=k\}}\bigr].
\end{align}
Here, $\Exp[e^{-s T_1}] = \lambda/(s+\lambda)$ is immediate since
$\Phi$ is a homogeneous Poisson process with intensity~$\lambda$.
In considering $\Exp_\Phi^0[e^{-s B_0}\,\ind{\{C_1=k\}}]$, we note
that there may be one or more pushed-out services in a busy period.
Let $M_0 = \min\{n=0,1,2,\ldots \mid S_n \le \tau_n\}$, which
represents the index of the first packet completed for service after
$T_0$.
Note that $M_0<\infty$ $\Prb_\Phi^0$-a.s.\ since
$\Prb_\Phi^0(S_0\le\tau_0) = \Lpl_S(\lambda)>0$.
Then, $\{M_0=n\} = \{S_i>\tau_i,\, i=0,1,\ldots,n-1; S_n\le\tau_n\}$
and $B_0 = \sum_{i=0}^{M_0-1}\tau_i + S_{M_0}$.
Since $(\tau_n,S_n)$, $n\in\Z$, are mutually independent and
identically distributed, we have
\begin{align}\label{eq:prf-marginalAoI4}
  \Exp_\Phi^0\bigl[e^{-s B_0}\,\ind{\{C_1=k\}}\bigr]
  &= \sum_{n=0}^\infty
       \Exp_\Phi^0\bigl[
         e^{-s B_0}\,\ind{\{C_1=k\}}\,\ind{\{M_0=n\}}
       \bigr]
  \nonumber\\
  &= \Exp_\Phi^0\bigl[
       e^{-s S_0}\,\ind{\{S_0 \le \tau_0\}}\,\ind{\{c_0 = k\}}
     \bigr]
     \sum_{n=0}^\infty
       \bigl(
         \Exp_\Phi^0\bigl[
           e^{-s \tau_0}\,\ind{\{S_0 > \tau_0\}}
         \bigr]
       \bigr)^n.  
\end{align}
Here, a similar way to obtaining \eqref{eq:prf-marginalAoI1} leads to
\[
  \Exp_\Phi^0\bigl[
    e^{-s S_0}\,\ind{\{S_0 \le \tau_0\}}\,\ind{\{c_0 = k\}}
  \bigr]
  = \frac{\lambda_k}{\lambda}\,\Lpl_{S,k}(s+\lambda),
\]
and on the other hand,
\begin{align*}\label{eq:prf-marginalAoI6}
  \Exp_\Phi^0\bigl[
    e^{-s \tau_0}\,\ind{\{S_0 > \tau_0\}}
  \bigr]
  &= \Exp_\Phi^0\bigl[\Exp_\Phi^0\bigl[
         e^{-s \tau_0}\,\ind{\{S_0 > \tau_0\}} \mid S_0
     \bigr]\bigr]
  \nonumber\\
  &= \lambda\,\Exp_\Phi^0\biggl[
       \int_0^{S_0} e^{-(s+\lambda)\,x}\,\dd x
     \biggr]
  \nonumber\\
  &= \frac{\lambda\,\bigl(1 - \Lpl_S(s+\lambda)\bigr)}
          {s+\lambda}.
\end{align*}
Plugging these into \eqref{eq:prf-marginalAoI4}, and then to
\eqref{eq:prf-marginalAoI3}, we obtain
\begin{equation}\label{eq:prf-marginalAoI7}
  \Exp_\Psi^0\bigl[e^{-s U_1}\,\ind{\{C_1=k\}}\bigr]
  = \frac{\lambda_k\,\Lpl_{S,k}(s+\lambda)}
         {s + \lambda\,\Lpl_S(s+\lambda)}.
\end{equation}
The same discussion as above except for replacing $\{k\}$ by
$\K\setminus\{k\}$ derives
\begin{equation}\label{eq:prf-marginalAoI8}
  \Exp_\Psi^0\bigl[
    e^{-s U_1}\,\ind{\{C_1 \in \K\setminus\{k\}\}}
  \bigr]
  = \frac{\lambda_{\K\setminus\{k\}}\,\Lpl_{S,\K\setminus\{k\}}(s+\lambda)}
         {s + \lambda\,\Lpl_S(s+\lambda)},
\end{equation}
and further plugging \eqref{eq:prf-marginalAoI7} and
\eqref{eq:prf-marginalAoI8} into \eqref{eq:prf-marginalAoI2}, we have
\begin{equation}\label{eq:prf-marginalAoI9}
  \Exp_\Psi^0[e^{-s U_{k,1}}]
  = \frac{\lambda_k\,\Lpl_{S,k}(s+\lambda)}
         {s + \lambda_k\,\Lpl_{S,k}(s+\lambda)}.
\end{equation}
Finally, substitution of \eqref{eq:prf-marginalAoI1} and
\eqref{eq:prf-marginalAoI9} into \eqref{eq:prf-marginalAoI0} yields
\eqref{eq:marginalAoI}.
\end{IEEEproof}

We can obtain the marginal moments of $A_k(0)$, $k\in\K$, from
\eqref{eq:marginalAoI} in any order as far as they exist.
For example, the first two moments are given by
\begin{align}
  \Exp[A_k(0)]
  &= - \frac{\dd\,\Lpl_{A_k}(s)}{\dd s}
      \biggr|_{s=0}
   = \frac{1}{\lambda_k\,\Lpl_{S,k}(\lambda)},
  \label{eq:MeanAoI}\\
  \Exp[A_k(0)^2]
  &= \frac{\dd^2\,\Lpl_{A_k}(s)}{\dd s^2}
     \biggr|_{s=0}
   = \frac{2\bigl(1 + \lambda_k\,\Lpl_{S,k}^{(1)}(\lambda)\bigr)}
     {\lambda_k^2\,\Lpl_{S,k}(\lambda)^2},
  \nonumber
\end{align}
where $\Lpl_{S,k}^{(m)}$ denotes the $m$th derivative of $\Lpl_{S,k}$.
The variance and the coefficient of variation are then given by
\begin{align}
  \Var[A_k(0)]
  &= \frac{1 + 2\lambda_k\,\Lpl_{S,k}^{(1)}(\lambda)}
          {{\lambda_k}^2\,\Lpl_{S,k}(\lambda)^2},
  \label{eq:VarAoI}\\
  \CV(A_k(0))
  &= \sqrt{1 + 2\lambda_k\,\Lpl_{S,k}^{(1)}(\lambda)},
  \nonumber
\end{align}
where the numerator of \eqref{eq:VarAoI} is definitely nonnegative
since $1-ax\,e^{-bx}\ge0$ whenever $a>0$, $b>0$ and $a/b\le e$ (in our
case, $2\lambda_k \le e\,\lambda$ always holds).

\begin{remark}
When $K=1$, our system corresponds to the one considered in
\cite[Sec.~4]{KesiKonsZaza19} and indeed \eqref{eq:marginalAoI} and
\eqref{eq:MeanAoI} are the same as those presented in Corollary~4
in~\cite{KesiKonsZaza19}.
Furthermore, when the service time distributions are common to all
sources, our system is reduced to the one considered in
\cite{NajmTela18} and \eqref{eq:MeanAoI} becomes equal to
\cite[eq.~(8)]{NajmTela18}  when $\Lpl_{S,k}=\Lpl_S$ for all $k\in\K$.
In addition, it also agrees with \cite[eq.~(47)]{KaulYateGrut12} and
\cite[Theorem~2~(a)]{YateKaul19} when the service time distributions
are a common exponential one with mean $\mu^{-1}$; that is,
substituting $\Lpl_{S,k}(\lambda) = \Lpl_S(\lambda) = (1+\rho)^{-1}$
with $\rho = \lambda/\mu = \sum_{k=1}^K\lambda_k/\mu$, we have
$\Exp[A_k(0)] = (1 + \rho)/\lambda_k$, $k\in\K$.
\end{remark}

We now tackle one of our main purposes in this paper; that is, we
derive a closed-form expression for the joint Laplace transform of the
AoIs.

\begin{theorem}\label{thm:main}
For the $K$-source M/G/1/1 pushout server described above, the joint
Laplace transform~$\Lpl_{\bsym{A}}$ of the stationary
AoIs~$A_1(0),\ldots,A_K(0)$ is given by
\begin{equation}\label{eq:main}
  \Lpl_{\bsym{A}}(\bsym{s})
  = \lambda_1\cdots\lambda_K
    \sum_{(j_1,\ldots,j_K)\in\sigma[\K]}
      \prod_{k=1}^K
        \frac{\Lpl_{S,j_k}(\Bar{s}_{j[k]} + \lambda)}
             {\Bar{s}_{j[k]} + \Bar{\lambda}_{j[k]}\,
                              \Lpl_{S,\,j[k]}(\Bar{s}_{j[k]} + \lambda)}, 
  \quad \bsym{s} = (s_1,\ldots,s_K) \in[0,\infty)^K,
\end{equation}
where $\sigma[\K]$ denotes the set of all permutations of $\K$ and
$j[k] = \{j_k, j_{k+1},\ldots,j_K\}$ for a permutation
$(j_1,\ldots,j_K) \in\sigma[\K]$.
\end{theorem}

\begin{IEEEproof}
We use \eqref{eq:AoI_joint_Lpl} in Theorem~\ref{thm:AoI_joint_Lpl}.
Due to the independent increments property of a Poisson process and
the independence of the service times, the behaviors of the server
before and after packet generations and also those before and after
service completions are independent.
Therefore, \eqref{eq:AoI_joint_Lpl} becomes
\begin{align}\label{eq:mg1-prf0}
  \Bar{s}\,\Lpl_{\bsym{A}}(\bsym{s})
  &= \lambda_\Psi\,
     \bigl( 1- \Exp_\Psi^0[e^{-\Bar{s} U_1}]\bigr)
  \nonumber\\
  &\quad\mbox{}\times   
     \sum_{(j_1,\,\ldots,\,j_K)\in\sigma[\K]}
       \prod_{k=1}^{K-1}
         \Exp_\Psi^0\bigl[
           \exp\bigl\{
             - s_{j_k} D_{j_k,\,0}
             - \Bar{s}_{j[k+1]} (U_{j_k,\,0} - U_{j_{k+1},\,0})
           \bigr\}\,
           \ind{\{\eta_k=j_k\}}
         \bigr]
  \nonumber\\
  &\quad\mbox{}\times
       \Exp_\Psi^0\bigl[  
         \exp\bigl( - s_{j_K} D_{j_K,\,0}\bigr)\,
         \ind{\{\eta_K=j_K\}}
       \bigr].
\end{align}
For $\Exp_\Psi^0[e^{-\Bar{s} U_1}]$ above, the same discussion as
obtaining \eqref{eq:prf-marginalAoI7} shows
\begin{equation}\label{eq:Lpl_U}
  \Exp_\Psi^0[e^{-\Bar{s} U_1}]
  = \frac{\lambda\,\Lpl_S(\Bar{s} + \lambda)}
         {\Bar{s} + \lambda\,\Lpl_S(\Bar{s} + \lambda)}.
\end{equation}
We next consider the last term~$\Exp_\Psi^0\bigl[\exp\bigl( -
  s_{j_K} D_{j_K,\,0}\bigr)\, \ind{\{\eta_K=j_K\}}\bigr]$ in
\eqref{eq:mg1-prf0}.
The stationarity of $\{U_{n+1}-U_n\}_{n\in\Z}$ in $\Prb_\Psi^0$
enables us to use a similar discussion to obtaining
\eqref{eq:prf-marginalAoI1} and 
\begin{align}\label{eq:last_term0}
 \Exp_\Psi^0\bigl[
   \exp\bigl( - s_{j_K} D_{j_K,0}\bigr)\,
   \ind{\{\eta_K=j_K\}}
 \bigr]
 &= \frac{\lambda_{\Psi_{j_K}}}{\lambda_\Psi}\,
    \Lpl_{D_{j_K}}(s_{j_K})
  \nonumber\\
  &= \frac{\lambda_{j_K}\,\Lpl_{S,\,j_K}(s_{j_K} + \lambda)}
          {\lambda\,\Lpl_S(\lambda)},
\end{align}
where we use $\lambda_\Psi = \lambda\,\Lpl_S(\lambda)$.
Thus, it remains to solve
\[
  \Exp_\Psi^0\bigl[
    \exp\bigl\{
      - s_{j_k} D_{j_k,0}
      - \Bar{s}_{j[k+1]} (U_{j_k,0} - U_{j_{k+1},0})
    \bigr\}\,
    \ind{\{\eta_k=j_k\}}
  \bigr].
\]
Similar to considering \eqref{eq:prf-marginalAoI2}, there may be one
or more service completions during $(U_{j_{k+1},0}, U_{j_k,0})$, but
if any, they must be of the sources $j_1,j_2,\ldots,j_k$ by the
definition of $\eta_k$, $k\in\K$.
Therefore, since the server is always reset at $U_n$, $n\in\Z$, the
above term is equal to
\begin{align}\label{eq:3rd_term0}
  &\sum_{m=0}^\infty
       \Exp_\Psi^0\Bigl[
         \exp\bigl\{
           - s_{j_k} D_{j_k,0}
           - \Bar{s}_{j[k+1]} (U_{j_k,0} - U_{j_{k+1},0})
         \bigr\}\,
         \ind{\{\eta_k=j_k\}}\,
         \ind{\{\Psi(U_{j_{k+1},0},U_{j_k,0}) = m\}}
       \Bigr]
  \nonumber\\
  &= \Exp_\Psi^0\Bigl[
       \exp\bigl\{
         - s_{j_k} D_1 - \Bar{s}_{j[k+1]} U_1
       \bigr\}\,
       \ind{\{C_1 = j_k\}}
     \Bigr]
     \sum_{m=0}^\infty
       \Bigl(
         \Exp_\Psi^0\bigl[
           \exp\bigl\{- \Bar{s}_{j[k+1]} U_1\bigr\}\,
           \ind{\{C_1\in \UBar{j}[k]\}}
         \bigr]
       \Bigr)^m,
\end{align}
where $\UBar{j}[k] = \{j_1,\ldots,j_k\} = \K\setminus j[k+1]$.
Similar to obtaining \eqref{eq:prf-marginalAoI7} and
\eqref{eq:prf-marginalAoI8}, we have
\begin{align*}\label{eq:3rd_term2}
  &\Exp_\Psi^0\Bigl[
     \exp\Bigl\{
       - s_{j_k} D_1 - \Bar{s}_{j[k+1]} U_1
     \Bigr\}\,
     \ind{\{C_1 = j_k\}}  
   \Bigr]
  \nonumber\\
  &= \Exp\bigl[e^{-\Bar{s}_{j[k+1]} T_1}\bigr]
     \sum_{n=0}^\infty
       \Bigl(
         \Exp_\Phi^0\bigl[
           e^{-\Bar{s}_{j[k+1]} \tau_0}\,
           \ind{\{S_0 > \tau_0\}}
         \bigr]
       \Bigr)^n
     \Exp_\Phi^0\bigl[
       e^{-\Bar{s}_{j[k]} S_0}\,
       \ind{\{S_0\le \tau_0\}}\,\ind{\{c_0 = j_k\}}
     \bigr]
  \nonumber\\
  &= \frac{\lambda_{j_k}\,\Lpl_{S,j_k}(\Bar{s}_{j[k]} + \lambda)}
         {\Bar{s}_{j[k+1]} + \lambda\,\Lpl_S(\Bar{s}_{j[k+1]} + \lambda)},
\end{align*}
where we note that $\Bar{s}_{j[k]} = s_{j_k} + \Bar{s}_{j[k+1]}$, and
\[
  \Exp_\Psi^0\bigl[
    \exp\bigl\{- \Bar{s}_{j[k+1]} U_1\bigr\}\,
    \ind{\{C_1\in \UBar{j}[k]\}}
  \bigr]
  = \frac{\lambda_{\UBar{j}[k]}\,\Lpl_{S,\UBar{j}[k]}(\Bar{s}_{j[k+1]} + \lambda)}
         {\Bar{s}_{j[k+1]} + \lambda\,\Lpl_S(\Bar{s}_{j[k+1]} + \lambda)}.
\]
Therefore, \eqref{eq:3rd_term0} amounts to
\begin{equation}\label{eq:3rd_term3}
  \text{\eqref{eq:3rd_term0}}
  = \frac{\lambda_{j_k}\,\Lpl_{S,j_k}(\Bar{s}_{j[k]} + \lambda)}
         {\Bar{s}_{j[k+1]} + \Bar{\lambda}_{j[k+1]}\,\Lpl_{S,\,j[k+1]}(\Bar{s}_{j[k+1]} +\lambda)}.
\end{equation}
Finally, plugging \eqref{eq:Lpl_U}, \eqref{eq:last_term0} and
\eqref{eq:3rd_term3} into \eqref{eq:mg1-prf0} and using $\lambda_\Psi
= \lambda\,\Lpl_S(\lambda)$, we obtain \eqref{eq:main}.
\end{IEEEproof}

\begin{remark}
Unfortunately, it is complicated to show that \eqref{eq:main} with
$s_j=0$ for $j\in\K\setminus\{k\}$ agrees with \eqref{eq:marginalAoI}
except for the case of small~$K$.
We thus focus on the two-source system in the next subsection.
\end{remark}

\subsection{Correlation coefficient in the two-source
  system}\label{subsec:cc}

In this subsection, we investigate the correlation coefficient of AoIs
in the two-source system.
When $K=2$, \eqref{eq:main} in Theorem~\ref{thm:main} is reduced to
\begin{equation}\label{eq:JointLpl2}
  \Lpl_{\bsym{A}}(s_1, s_2)
  = \frac{\lambda_1\,\lambda_2}
         {\Bar{s} + \lambda\,\Lpl_S(\Bar{s}+\lambda)}
    \sum_{k=1}^2
      \frac{\Lpl_{S,k}(s_k+\lambda)\,\Lpl_{S,3-k}(\Bar{s}+\lambda)}
           {s_k + \lambda_k\,\Lpl_{S,k}(s_k+\lambda)},
  \quad s_1\ge0,\: s_2\ge0.
\end{equation}
In this case, we can easily confirm that both $\Lpl_{\bsym{A}}(s,0)$
and $\Lpl_{\bsym{A}}(0,s)$ agree with \eqref{eq:marginalAoI}.
The expectation of the product~$A_1(0)\,A_2(0)$ is obtained from
\eqref{eq:JointLpl2} as
\begin{align*}
  \Exp\bigl[
    A_1(0)\,A_2(0)
  \bigr]
  &= \frac{\partial^2}{\partial s_1\,\partial s_2}
       \Lpl_{\bsym{A}}(s_1, s_2)
     \biggr|_{s_1=s_2=0}
  \\
  &= \frac{1}{\lambda\,\Lpl_S(\lambda)}
     \sum_{k=1}^2\frac{\Lpl_{S,k}^{(1)}(\lambda)}{\Lpl_{S,k}(\lambda)}
     + \prod_{k=1}^2\frac{1}{\lambda_k\,\Lpl_{S,k}(\lambda)}.
\end{align*}
Therefore, combining this with \eqref{eq:MeanAoI}, we have the
covariance of $A_1(0)$ and $A_2(0)$ as
\[
  \Cov\bigl(A_1(0), A_2(0)\bigr)
  = \frac{1}{\lambda\,\Lpl_S(\lambda)}
  \sum_{k=1}^2\frac{\Lpl_{S,k}^{(1)}(\lambda)}{\Lpl_{S,k}(\lambda)},
\]
from which we can see that $A_1(0)$ and $A_2(0)$ are negatively
correlated since the first derivative of the Laplace transform for a
nonnegative random variable is always nonpositive.
Further combination with \eqref{eq:VarAoI} gives the correlation
coefficient;
\begin{equation}\label{eq:CC}
  \CC(A_1(0), A_2(0))
  = \frac{\lambda_1\,\lambda_2\bigl(
            \Lpl_{S,1}^{(1)}(\lambda)\,\Lpl_{S,2}(\lambda)
            + \Lpl_{S,1}(\lambda)\,\Lpl_{S,2}^{(1)}(\lambda)
          \bigr)}
         {\lambda\,\Lpl_S(\lambda)\sqrt{
            \bigl(1 + 2\lambda_1\,\Lpl_{S,1}^{(1)}(\lambda)\bigr)
            \bigl(1 + 2\lambda_2\,\Lpl_{S,2}^{(1)}(\lambda)\bigr) 
          }}.                    
\end{equation}

We note that when the service time distributions are a common
exponential one with mean $\mu^{-1}$; that is, $\Lpl_{S,k}(s) =
\Lpl_S(s) = (1+s/\mu)^{-1}$ for $k=1,2$, \eqref{eq:CC} is indeed
reduced to that obtained in~\cite{JianTokuWadaYaji20}.
Some properties of the correlation coefficient are collected in the
following proposition.

\begin{proposition}\label{prp:cc}
\begin{enumerate}  
\item\label{prp:cc1} For $k=1,2$,
  $\lim_{\lambda_k\downarrow0}\CC(A_1(0),A_2(0)) = 0$ and, if
  $\Lpl_{S,3-k}(s) = O(\Lpl_{S,k}(s))$ as $s\to\infty$, then
  $\lim_{\lambda_k\to\infty}\CC(A_1(0),A_2(0)) = 0$.

\item\label{prp:cc2} Suppose that we can choose the service time
  distribution under the constraint that $\Lpl_{S,1} = \Lpl_{S,2}$ and
  $\lambda=\lambda_1+\lambda_2$ is fixed.
  Then, $\CC(A_1(0),A_2(0))$ takes the minimum value when the service
  times are deterministic and equal to $\lambda^{-1}$.
  Furthermore, this minimum value is bounded below by
  $-\bigl(2(e-1)\bigr)^{-1} \approx -0.290988$, which is realized if
  and only if $\lambda_1=\lambda_2 = \lambda/2$.

\item\label{prp:cc3} When the service times follow a common gamma
  distribution with shape parameter~$\alpha>0$ and rate
  parameter~$\mu>0$ (that is, the mean service time is $\alpha/\mu$)
  in both the sources, then $\lim_{\mu\downarrow0}\CC(A_1(0),A_2(0)) =
  \lim_{\mu\to\infty}\CC(A_1(0),A_2(0)) = 0$.

\item\label{prp:cc4} Suppose that $\lambda=\lambda_1+\lambda_2$ is
  fixed and that the service times follow a common gamma distribution
  with shape parameter~$\alpha>0$ and rate parameter~$\mu>0$.
  Then, $\CC(A_1(0),A_2(0))$ takes the minimum value when
  $\mu=\alpha\lambda$.
  Furthermore, this minimum value is bounded below by
  $-\bigl[2\bigl((1 + 1/\alpha)^{\alpha+1} - 1\bigr)\bigr]^{-1}$,
  which is realized if and only if $\lambda_1=\lambda_2 = \lambda/2$.
\end{enumerate}
\end{proposition}

\begin{IEEEproof}
\ref{prp:cc1})
The convergence as $\lambda_k\downarrow0$ is immediate from
\eqref{eq:CC}.
For the convergence as $\lambda_k\to\infty$, it is sufficient by
symmetry to show that $\lim_{\lambda_1\to\infty}\CC(A_1(0),A_2(0)) =
0$ when $\Lpl_{S,2}(s) = O(\Lpl_{S,1}(s))$ as $s\to\infty$.
Dividing the numerator and the denominator on the right-hand side of
\eqref{eq:CC} by $\lambda_1\,\Lpl_{S,1}(\lambda)$, we have
\begin{equation}\label{eq:CC0}
  \CC(A_1(0),A_2(0))
  = \frac{\lambda_2\biggl(
            \Lpl_{S,1}^{(1)}(\lambda)\,
            \dfrac{\Lpl_{S,2}(\lambda)}{\Lpl_{S,1}(\lambda)}
            + \Lpl_{S,2}^{(1)}(\lambda)
          \biggr)}
         {\biggl(
            1 + \dfrac{\lambda_2\,\Lpl_{S,2}(\lambda)}
                      {\lambda_1\,\Lpl_{S,1}(\lambda)}
          \biggr)\sqrt{
            \bigl(1 + 2\lambda_1\,\Lpl_{S,1}^{(1)}(\lambda)\bigr)
            \bigl(1 + 2\lambda_2\,\Lpl_{S,2}^{(1)}(\lambda)\bigr) 
          }}.                    
\end{equation}
Since $S_0$ is nonnegative, $e^{-s S_0}\le 1$ and $s\,S_0\,e^{-s
  S_0}\le e^{-1}$ for $s\ge0$.
Therefore, the dominated convergence theorem leads to
$\Lpl_{S,k}(s)\to0$ and $s\,\Lpl_{S,k}^{(1)}(s)\to0$ as $s\to\infty$
(and of course, $\Lpl_{S,k}^{(1)}(s)\to0$ as $s\to\infty$).
Hence, the numerator on the right-hand side of \eqref{eq:CC0} goes to
zero as $\lambda_1\to\infty$ when $\Lpl_{S,2}(s) = O(\Lpl_{S,1}(s))$
while the denominator goes to one.

\ref{prp:cc2})
When $\Lpl_{S,1}=\Lpl_{S,2}=\Lpl_S$, we have from \eqref{eq:CC} that
\begin{align*}
  -\frac{1}{\CC(A_1(0),A_2(0))}
  &= \frac{\lambda}{2\lambda_1\lambda_2}
     \sqrt{%
       \biggl(-\frac{1}{\Lpl_S^{(1)}(\lambda)}-2\lambda_1\biggr)
       \biggl(-\frac{1}{\Lpl_S^{(1)}(\lambda)}-2\lambda_2\biggr)%
     }
  \nonumber\\
  &\ge \frac{\lambda}{2\lambda_1\lambda_2}
       \sqrt{(e\lambda - 2\lambda_1)(e\lambda-2\lambda_2)},
\end{align*}
where the inequality follows from $-\Lpl_S^{(1)}(\lambda)\le
(e\lambda)^{-1}$ and the equality holds only when the service times
are deterministic and equal to $\lambda^{-1}$.
Furthermore, applying the inequality of arithmetic and geometric means
$\lambda/2 = (\lambda_1+\lambda_2)/2 \ge \sqrt{\lambda_1\lambda_2}$
twice, the last expression above is bounded below by
\begin{align*}
  \frac{\lambda}{2\lambda_1\lambda_2}
  \sqrt{(e\lambda - 2\lambda_1)(e\lambda-2\lambda_2)}
  &\ge \sqrt{
         e(e-2)\frac{\lambda^2}{\lambda_1\lambda_2} + 4
       }
  \nonumber\\
  &\ge 2(e-1),
\end{align*}
where both the equalities hold if and only if $\lambda_1=\lambda_2$.

\ref{prp:cc3})
Since $\Lpl_{S,1}(s) = \Lpl_{S,2}(s) = \Lpl_S(s) =
(1+s/\mu)^{-\alpha}$, \eqref{eq:CC} is reduced to
\begin{equation}\label{eq:-1/CC}
  \CC(A_1(0),A_2(0))
  = - \frac{2\alpha\lambda_1\lambda_2}
           {\lambda\sqrt{
              \bigl(
                (\mu+\lambda)^{\alpha+1}/\mu^\alpha - 2\alpha\lambda_1
              \bigr)\bigl(
                (\mu+\lambda)^{\alpha+1}/\mu^\alpha - 2\alpha\lambda_2
              \bigr)
           }},
\end{equation}
which clearly goes to zero as $\mu\downarrow0$.
Furthermore, the above expression is also rewritten as
\[
  \CC(A_1(0),A_2(0))
  = - \frac{2\alpha\lambda_1\lambda_2/\mu}
           {\lambda\sqrt{
              \bigl(
                (1+\lambda/\mu)^{\alpha+1} - 2\alpha\lambda_1/\mu 
              \bigr)\bigl(
                (1+\lambda/\mu)^{\alpha+1} - 2\alpha\lambda_2/\mu
              \bigr)
           }},
\]
which is confirmed to go to zero as $\mu\to\infty$.

\ref{prp:cc4})
From~\eqref{eq:-1/CC}, we have
\begin{align*}
  -\frac{1}{\CC(A_1(0),A_2(0))}
  &= \frac{\lambda}{2\lambda_1\lambda_2}
     \sqrt{
       \biggl(
         \frac{(\mu+\lambda)^{\alpha+1}}{\alpha\mu^\alpha}
         - 2\lambda_1
       \biggr)
       \biggl(
         \frac{(\mu+\lambda)^{\alpha+1}}{\alpha\mu^\alpha}
         - 2\lambda_2
       \biggr)
     }
  \\
  &\ge \frac{\lambda}{2\lambda_1\lambda_2}
       \sqrt{
         \biggl(
           \lambda\Bigl(1+\frac{1}{\alpha}\Bigr)^{\alpha+1}
           - 2\lambda_1
         \biggr)
         \biggl(
           \lambda\Bigl(1+\frac{1}{\alpha}\Bigr)^{\alpha+1}
           - 2\lambda_2
         \biggr)
       },    
\end{align*}
where the inequality follows from $(\mu+\lambda)^{\alpha+1}/\mu^\alpha
\ge \alpha\lambda(1+1/\alpha)^{\alpha+1}$ and the equality holds when
$\mu=\alpha\lambda$.
Furthermore, applying the inequality of arithmetic and geometric means
$\lambda/2 = (\lambda_1+\lambda_2)/2 \ge \sqrt{\lambda_1\lambda_2}$,
we can see that the last expression above is bounded below by
$2\bigl((1+1/\alpha)^{\alpha+1}-1\bigr)$ and this lower bound is
realized if and only if $\lambda_1=\lambda_2$.
\end{IEEEproof}

\section{Numerical experiments}\label{sec:experiments}

\begin{figure*}[!t]
\centering%
\subfloat[$\lambda_1=1.0$, {$1/\Exp_\Phi^0[S_0]=6.0$}]%
  {\includegraphics[width=.48\linewidth]{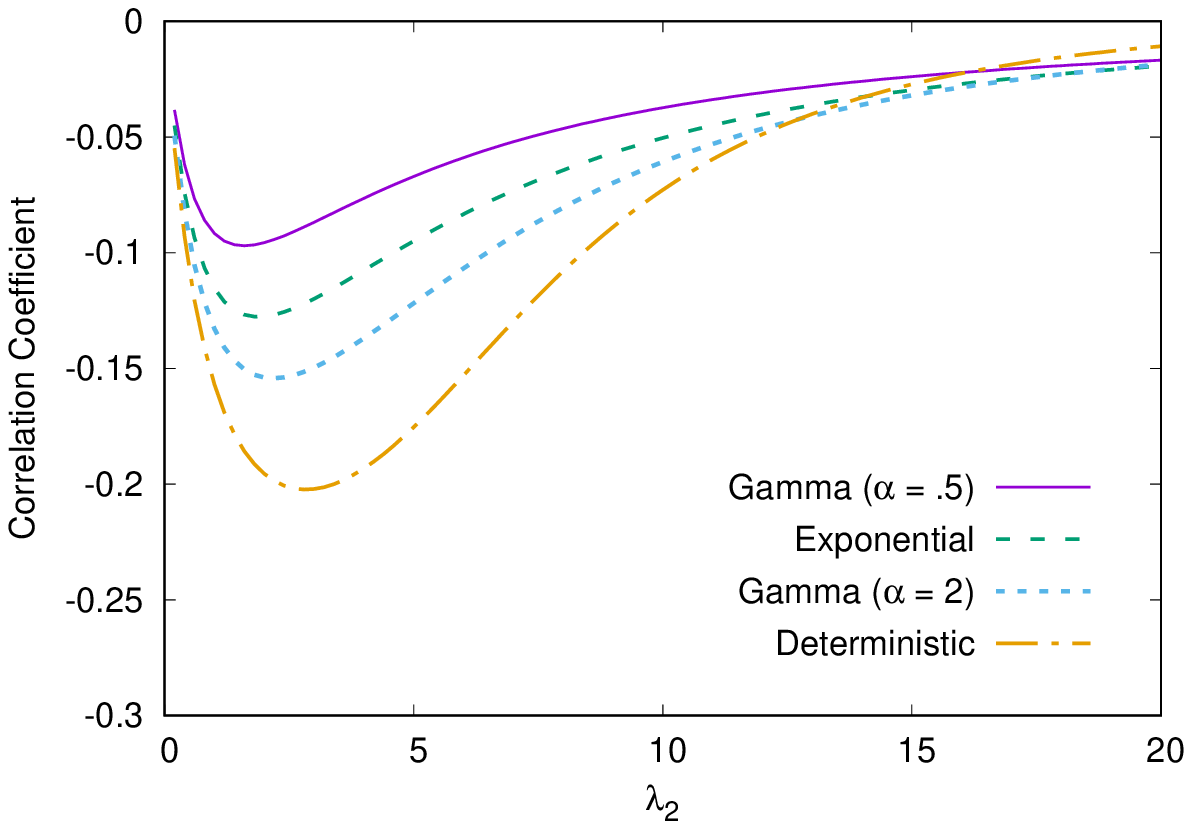}%
   \label{fig:lambda2-change-16}}%
\hfil
\subfloat[$\lambda_1=3.0$, {$1/\Exp_\Phi^0[S_0]=6.0$}]%
  {\includegraphics[width=.48\linewidth]{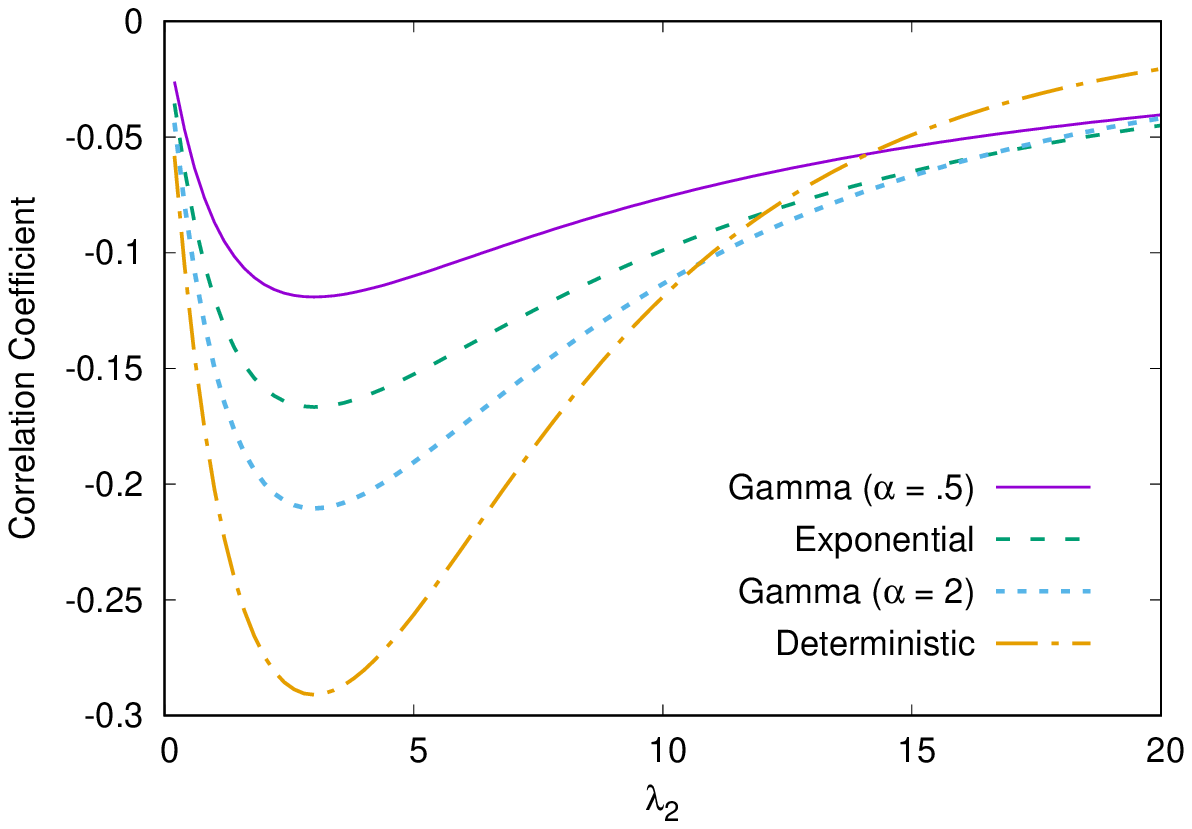}%
   \label{fig:lambda2-change-36}}
\caption{Values of correlation coefficient of AoIs in the two-source
  M/G/1/1 pushout server for different values of $\lambda_2$.}%
\label{fig:lambda2-change}
\end{figure*}

We here confirm the properties of the correlation coefficient of AoIs
proved in Proposition~\ref{prp:cc} through numerical experiments.
Throughout the experiments, we use the two-source M/G/1/1 pushout
server with a common service time distribution.
Figure~\ref{fig:lambda2-change} plots the values of the correlation
coefficient for different values of $\lambda_2$ when the values of
$\lambda_1$ and $\Exp_\Phi^0[S_0]$ are fixed, where
Figure~\ref{fig:lambda2-change-16}) and~\ref{fig:lambda2-change-36})
show the cases of $(\lambda_1, 1/\Exp_\Phi^0[S_0]) =  (1.0, 6.0)$ and
$(\lambda_1, 1/\Exp_\Phi^0[S_0]) = (3.0, 6.0)$, respectively.
Note that $1/\Exp_\Phi^0[S_0] = \mu/\alpha$ when the service times
are gamma distributed with shape parameter~$\alpha>0$ and rate
parameter~$\mu>0$.
From these figures, we can see property~\ref{prp:cc1}) in
Proposition~\ref{prp:cc}; that is, the correlation coefficient goes to
zero as $\lambda_2\downarrow0$ and as $\lambda_2\to\infty$.
As another property, we remark that the correlation coefficient takes
the minimum value at $\lambda_2 = 1/\Exp_\Phi^0[S_0] - \lambda_1$ in
the case of Figure~\ref{fig:lambda2-change-36}), but it does not in
the case of \ref{fig:lambda2-change-16}).
It does not contradict properties~\ref{prp:cc2}) and~\ref{prp:cc4}) in
Proposition~\ref{prp:cc}; that is, when $\lambda_1$ and $\lambda_2$
are given, the correlation coefficient takes the minimum value when
$1/\Exp_\Phi^0[S_0] = \lambda = \lambda_1 + \lambda_2$, but when
$\lambda_1$ and $\Exp_\Phi^0[S_0]$ are given, it does not always take
the minimum value at either $\lambda_2 = 1/\Exp_\Phi^0[S_0] -
\lambda_1$ or $\lambda_2=\lambda_1$.

\begin{figure*}[!t]
\centering
\subfloat[$\lambda_1=\lambda_2=3.0$]%
  {\includegraphics[width=.48\linewidth]{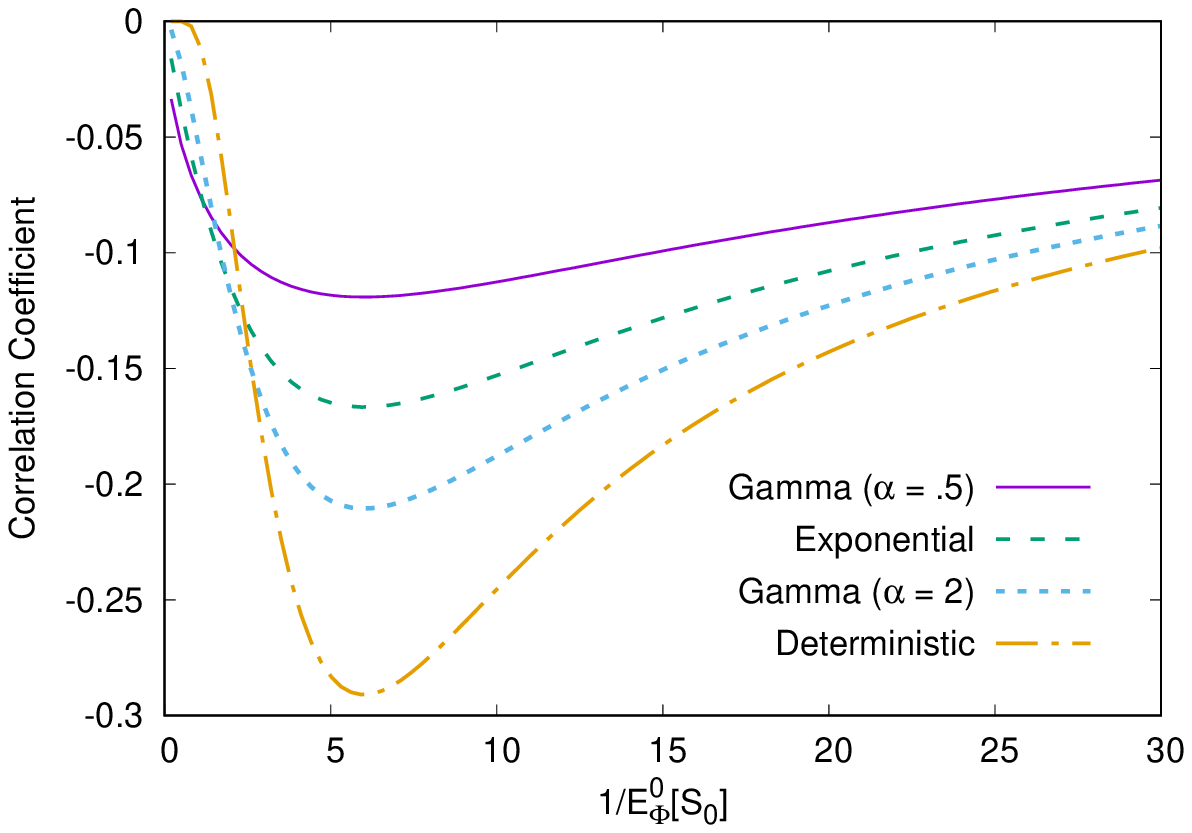}%
   \label{fig:mu-change33}}%
\hfil    
\subfloat[$\lambda_1=1.0$, $\lambda_2=5.0$]%
  {\includegraphics[width=.48\linewidth]{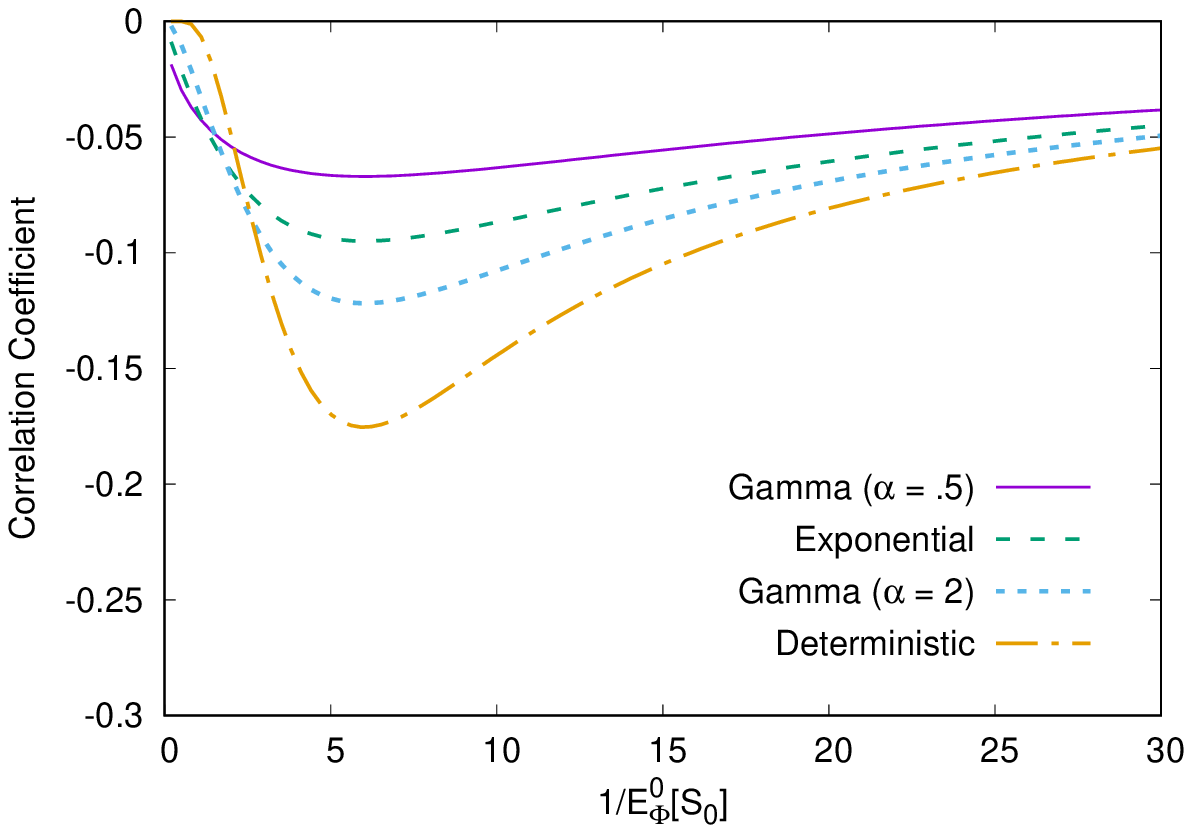}%
   \label{fig:mu-change15}}%
\caption{Values of correlation coefficient of AoIs in the two-source
  M/G/1/1 pushout server for different values of
  $1/\Exp_\Phi^0[S_0]$.}%
\label{fig:mu-change}
\end{figure*}

Figure~\ref{fig:mu-change} plots the values of the correlation
coefficient of AoIs for different values of $1/\Exp_\Phi^0[S_0]$
when $\lambda_1$ and $\lambda_2$ are fixed.
Figure~\ref{fig:mu-change33}) and~\ref{fig:mu-change15}) show
respectively the cases of $\lambda_1 = \lambda_2$ and
$\lambda_1\ne\lambda_2$ while the value of
$\lambda=\lambda_1+\lambda_2$ remains the same in both the figures.
From these figures, we can observe the properties~\ref{prp:cc2}),
\ref{prp:cc3}) and \ref{prp:cc4}) in Proposition~\ref{prp:cc}; that
is, the correlation coefficient goes to zero as
$1/\Exp_\Phi^0[S_0]\downarrow0$ and as $1/\Exp_\Phi^0[S_0]\to\infty$;
when $\lambda=\lambda_1+\lambda_2$ is fixed, the correlation
coefficient takes the minimum value when the service times are
deterministic and equal to $\lambda^{-1}$, in addition, when the
service times are gamma distributed, it takes the minimum value when
$1/\Exp_\Phi^0[S_0]=\lambda$, where these minimum values are further
minimized when $\lambda_1=\lambda_2 = \lambda/2$.

\section{Conclusion}\label{sec:conclusion}

In this paper, we have considered multi-source status update systems
and have provided a framework to investigate the joint performance of
multiple AoIs, which are defined for the individual sources.
Specifically, we have derived a general formula satisfied by the joint
Laplace transform of the stationary AoIs.
Then, we have applied this formula to a multi-source pushout server and
have shown a closed-form formula of the joint Laplace transform of
the AoIs in the case with independent M/G inputs.
Furthermore, we have revealed some properties of the correlation
coefficient of AoIs in the two-source system.
In the future, we expect that our general formula will be utilized to
evaluate the joint performance of AoIs in many multi-source status
update systems and will become useful for development of various systems in
the real world.

\appendices
\section{Proof of Lemma~\ref{lem:joint-stationary}}\label{sec:Prf_stationary}

We suppose that all random elements in this paper is defined on a
probability space~$(\Omega, \F, \Prb)$.
A family of shift operators~$\{\theta_t\}_{t\in\R}$ is defined on
$(\Omega,\F)$ such that $\theta_t$:~$\Omega\to\Omega$ is measurable
and bijective satisfying $\theta_s\circ\theta_t=\theta_{s+t}$ for $s$,
$t\in\R$, where $\theta_0$ is the identity; so that $\theta_t^{-1} =
\theta_{-t}$ for $t\in\R$.
The probability measure~$\Prb$ is assumed to be invariant to
$\{\theta_t\}_{t\in\R}$ (in other words, $\{\theta_t\}_{t\in\R}$
preserves $\Prb$) in the sense that $\Prb\circ\theta_t^{-1} = \Prb$
for $t\in\R$.
Then, we can assume that the marked point process~$\Psi_{C,D}$
satisfying Assumption~\ref{asm1} is compatible with
$\{\theta_t\}_{t\in\R}$ in the sense that $(U_n, C_n,
D_n)\circ\theta_t = (U_{\Psi(0,\,t]+n} - t, C_{\Psi(0,\,t]+n},
D_{\Psi(0,\,t]+n})$ for each $n\in\Z$ and $t\in\R$, where 
$\Psi(0,\,t] = -\Psi(t,\,0]$ for $t<0$ conventionally
  (see~\cite{BaccBrem03}).

In the setup above, the proof of the first assertion in
Lemma~\ref{lem:joint-stationary} is achieved by showing that, for each
$k\in\K$, the marked point process~$\Psi_{k,D}$ and the AoI
process~$\{A_k(t)\}_{t\in\R}$ are both compatible with
$\{\theta_t\}_{t\in\R}$.
We first confirm $\Psi_{k,D}$.
By \eqref{eq:sub-departure}, $\Psi_{k,D}$ satisfies
\begin{align*}
  \Psi_{k,D}(B\times E)
  &= \sum_{n\in\Z}\ind{B}(U_{k,n})\,\ind{E}(D_{k,n}).
  \\
  &= \sum_{n\in\Z}\ind{B}(U_n)\,\ind{E}(D_n)\,\ind{\{k\}}(C_n),
  \quad B\in\B(\R),\: E\in\B([0,\infty)).
\end{align*}
The compatibility of $\Psi_{C,D}$ with $\{\theta_t\}_{t\in\R}$ implies
that, for each $n\in\Z$ and $t\in\R$, there exists a unique $n'\in\Z$
such that $(U_n-t, C_n, D_n) = (U_{n'}, C_{n'}, D_{n'})\circ\theta_t$
and then
\begin{align*}
  \Psi_{k,D}((B+t)\times E)
  &= \sum_{n\in\Z}\ind{B+t}(U_n)\,\ind{E}(D_n)\,\ind{\{k\}}(C_n)
  \\
  &= \sum_{n'\in\Z}
       \bigl(
         \ind{B}(U_{n'})\,\ind{E}(D_{n'})\,\ind{\{k\}}(C_{n'})
       \bigr)\circ\theta_t
  \\
  &= \Psi_{k,D}(B\times E)\circ\theta_t,
\end{align*}
where $B+t = \{s+t\mid s\in B\}$ for $t\in\R$ and $B\in\B(\R)$; that
is, $\Psi_{k,D}$ is compatible with $\{\theta_t\}_{t\in\R}$.
Therefore, we can show from \eqref{eq:AoI} that $A_k(s+t) =
A_k(s)\circ\theta_t$ for any $s$, $t\in\R$ in a similar way; that is,
$\{A_k(t)\}_{t\in\R}$ is also compatible with $\{\theta_t\}_{t\in\R}$.

We next show the second assertion in Lemma~\ref{lem:joint-stationary}.
Since the events~$\{D_{k,0}<\infty\}$ and $\{U_{k,1}-U_{k,0}<\infty\}$
are $\Prb_{\Psi_k}^0$-a.s.\ and
$\{\theta_{U_{k,n}}\}_{n\in\Z}$-invariant under Assumption~\ref{asm1},
\eqref{eq:AoI} ensures that the event~$\{A_k(0)<\infty\}$  is
$\Prb_{\Psi_k}^0$-a.s.\ and $\{\theta_{U_{k,n}}\}_{n\in\Z}$-invariant
as well.
Hence, \cite[p.~51, Property 1.6.2]{BaccBrem03} says that $A_k(0)<\infty$
$\Prb$-a.s.

\section{Proof of Lemma~\ref{lem:input-output}}\label{sec:Prf_input-output}

First, the simplicity of $\Psi$ is inherited from $\Phi$.
We next check~\ref{asm1}-\ref{asm12}).
In the same setting as in Section~\ref{sec:Prf_stationary}, we can
assume that the marked point process~$\Phi_{c,S}$ satisfying
Assumption~\ref{asm2} is compatible with $\{\theta_t\}_{t\in\R}$ in
the sense that $(T_n,c_n,S_n)\circ\theta_t = (T_{\Phi(0,\,t]+n}-t,
  c_{\Phi(0,\,t]+n}, S_{\Phi(0,\,t]+n})$ for each $n\in\Z$ and
    $t\in\R$, where $\Phi(0,t] = -\Phi(t,0]$ for $t<0$.
Then, we can see from \eqref{eq:input-output} that
$\Psi_{C,D}((B+t)\times\{k\}\times E) = \Psi_{C,D}(B\times\{k\}\times
E)\circ\theta_t$ holds for any $t\in\R$, $k\in\K$, $B\in\B(\R)$ and
$E\in\B([0,\infty))$; that is, $\Psi_{C,D}$ is also compatible with
$\{\theta_t\}_{t\in\R}$.

We now confirm \ref{asm1}-\ref{asm13}) with \eqref{eq:lambda_Psi}.
Let $\chi(t)$ denote the indicator that $\chi(t)=1$ when the server is
occupied by a packet at time~$t$ and $\chi(t)=0$ otherwise.
Then, $\chi(t)$ satisfies
\[
  \chi(t) = \sum_{n\in\Z}\ind{[T_n,\, T_n+\tau_n\wedge S_n)}(t),
  \quad t\in\R,
\]
which shows that $\{\chi(t)\}_{t\in}$ is also jointly stationary with
$\Phi_{c,S}$.
Furthermore, $\{\chi(t)\}_{t\in\R}$ satisfies
\[
  \chi(1)
  = \chi(0) + \Phi(0,1] - \Psi(0,1]
    - \sum_{n\in\Z}\ind{(0,1]}(T_n)\,\ind{\{S_{n-1}>\tau_{n-1}\}},  
\]
where the last term on the right-hand side represents the number of
pushouts during $(0,1]$.
Taking the expectations on both the sides above, the stationarity of
$\{\chi(t)\}_{t\in\R}$ implies
\begin{align}\label{eq:lambda_Psi_prf}
  \lambda_\Psi
  &= \lambda - \Exp\biggl[
                 \sum_{n\in\Z}\ind{(0,1]}(T_n)\,\ind{\{S_{n-1}>\tau_{n-1}\}}
               \biggr]
  \nonumber\\  
  &= \lambda - \lambda\,\Prb_\Phi^0(S_0>\tau_0)
   = \lambda\,\Prb_\Phi^0(S_0\le \tau_0),
\end{align}
and \eqref{eq:lambda_Psi} holds.

For \ref{asm1}-\ref{asm14}) with \eqref{eq:P_Psi(C0=k)}, the Neveu's
exchange formula (see \cite[p.~21]{BaccBrem03}) shows that
\begin{align}\label{eq:P(C=k)}
  \lambda_\Psi\,\Prb_\Psi^0(C_0 = k)
  &= \lambda\,\Exp_\Phi^0\biggl[
       \sum_{n\in\Z}\ind{\{C_n=k\}}\ind{[0,\,T_1)}(U_n)
     \biggr]
  \nonumber\\
  &= \lambda\,\Prb_\Phi^0(c_0 = k,\, S_0 \le \tau_0),
\end{align}
where the second equality follows from the observation that there
exists at most one service completion during $[0,\, T_1)$ and it
occurs only when the packet generated at $0$ is completed for service
without interruption.
Equation~\eqref{eq:P_Psi(C0=k)} then follows from
\eqref{eq:lambda_Psi_prf} and \eqref{eq:P(C=k)}.

Finally, \ref{asm1}-\ref{asm15}) is immediate from Neveu's exchange
formula as
\begin{align*}
  \lambda_{\Psi_k}\,\Prb_{\Psi_k}^0(D_0 < \infty)
  &= \lambda_k\,\Exp_{\Phi_k}^0\biggl[
       \sum_{n\in\Z}\ind{\{D_n<\infty\}}\,
         \ind{[0,T_{k,1})}(U_{k,n})
     \biggr]
   = \lambda_k\,\Prb_{\Phi_k}^0(S_0 \le \tau_0),
\end{align*}
where $\{T_{k,n}\}_{n\in\Z}$ denotes the sub-sequence of
$\{T_n\}_{n\in\Z}$ corresponding to $\Phi_k$ satisfying $\cdots <
T_{k,0} \le 0 < T_{k,1} < \cdots$ and the second equality follows from
a similar observation to \eqref{eq:P(C=k)}.
Hence, \ref{asm1}-\ref{asm15}) holds since the
equality~$\lambda_{\Psi_k} = \lambda_\Psi\,\Prb_\Psi^0(C_0=k) =
\lambda_k\,\Prb_{\Phi_k}^0(S_0 \le \tau_0)$ is obtained by
\eqref{eq:lambda_Psi} and \eqref{eq:P_Psi(C0=k)}.

\bibliographystyle{IEEETran}

\end{document}